\documentclass{article}
\usepackage{amsfonts}
\usepackage{epsfig,amsmath}
\usepackage{natbib}

\newcommand{\bd}[1]{\ensuremath{\mbox{\boldmath $#1$}}}

\begin{document}

\title{Boundary detection in disease mapping studies.}

\author{Duncan Lee $^{1}$ and Richard Mitchell $^{2}$\\
$^{1}$School of Mathematics and Statistics, University of Glasgow, Glasgow, UK\\
$^{2}$Public Health, School of Medicine, University of Glasgow, Glasgow, UK}


\maketitle


\begin{abstract}
{In disease mapping, the aim is to estimate the spatial pattern in disease risk over an extended geographical region, so that areas with elevated risks can be identified. A Bayesian hierarchical approach is typically used to produce such maps, which models the risk surface with a set of spatially smooth random effects. However, in complex urban settings there are likely to be boundaries in the risk surface, which separate populations that are geographically adjacent but have very different risk profiles. Therefore this paper proposes an approach for detecting such risk boundaries, and tests its effectiveness by simulation. Finally, the model is applied to lung cancer incidence data in Greater Glasgow, Scotland, between 2001 and 2005.}

{Boundary detection; Disease mapping; Spatial correlation}
\end{abstract}

\section{Introduction}
Disease mapping is the area of statistics that quantifies the spatial pattern in disease risk over an extended geographical region, such as a city of country. The study region is partitioned into a number of small non-overlapping areal units, such as electoral wards, and only the total number of cases in each unit is available. The majority of disease risk maps are produced using Bayesian hierarchical models, utilising a vector of covariate risk factors and  a set of random effects. The latter model any overdispersion or spatial correlation in the disease data (after the covariate effects have been allowed for), which can be caused by the presence of unmeasured risk factors that also have a spatial structure. The random effects are often represented by a Conditional Autoregressive (CAR) prior, which forces the effects in two areas to be correlated if those areas share a common border. \\

The production of such maps  provide a number of benefits to public health professionals, including the ability to investigate the association between disease prevalence and suspected risk factors. More recently, disease maps have been used to identify \emph{boundaries} (or \emph{cliffs} or \emph{discontinuities}) in the risk surface, which separate geographically adjacent areas that have high and low risks. Such boundaries are likely to exist in  complex urban settings, where rich and poor communities can be separated by just a few metres. The detection of such boundaries has a number of benefits,  including the ability to detect the spatial extent of a cluster of high risk areas. It is also important economically, because it allows health resources to be targeted at communities with the highest disease risks. Furthermore, the spatial units at which the health and covariate data are available are typically designed for administrative purposes, and thus often do not delineate between distinct neighbourhoods (i.e. groups of people with the same social circumstances, culture and behaviour). Therefore, boundaries in disease risk surfaces may also correspond to boundaries between different neighbourhoods, which are of interest because \emph{``their locations reflect underlying biological, physical, and/or social processes''} (\cite{jacquez2000}).\\

The statistical detection of boundaries in disease maps is also known as \emph{Wombling}, following the seminal article by \cite{womble1951}. More recently, a number of different approaches to this problem have been proposed, including the calculation of local statistics (\cite{boots2001}), and  a Bayesian random effects model (\cite{lu2005}).
In this paper we also use a Bayesian random effects model, and identify boundaries by measuring the level of dissimilarity between the populations living in neighbouring areas. We believe that risk boundaries are more likely to occur between populations that are very different, because homogeneous populations should have similar disease risks. The remainder of this paper is organised as follows. Section 2 provides a review of disease mapping and boundary detection methods, while Section 3 presents our proposed methodological extension. Section 4 assesses the efficacy of our approach using simulation, while Section 5 identifies boundaries in the risk surface for lung cancer cases in Greater Glasgow. Finally, Section 6 contains a concluding discussion and outlines future developments.

\section{Background}

\subsection{Disease mapping}
The data used to quantify disease risk are denoted by $\mathbf{y}=(y_{1},\ldots, y_{n})$ and  $\mathbf{E}=(E_{1},\ldots, E_{n})$, the former being the numbers of disease cases observed in each of $n$ non-overlapping areas within a specified time frame. The latter are the expected numbers of disease cases, which depend on the size and demographic structure of the population living within each area. Disease risk can be summarised by the standardised incidence ratio (SIR), which for area $k$ is given by $\hat{R}_{k}=y_{k}/E_{k}$.  Values above one represent areas with elevated risks of disease, while values less than one correspond to relatively healthy areas. However, elevated SIR values can occur by chance in areas where $E_{k}$ is small, so the set of disease risks for all $n$ areas are more commonly estimated using a Bayesian hierarchical model.  A general specification has been described by  \cite{elliott2000}, \cite{banerjee2004}, \cite{wakefield2007} and \cite{lawson2008},  and is given by

\begin{eqnarray}
Y_{k}|E_{k}, R_{k}&\sim&\mbox{Poisson}(E_{k}R_{k})~~~~\mbox{for }k=1,\ldots,n,\nonumber\\
\ln(R_{k})&=&\mathbf{x}_{k}^{\footnotesize\mbox{T}\normalsize}\bd{\beta}+\phi_{k}.\label{equation diseasemap}
\end{eqnarray}

Disease risk is modelled by covariates $\mathbf{x}_{k}^{\footnotesize\mbox{T}\normalsize}=(x_{k1},\dots,x_{kp})$, and random effects $\bd{\phi}=(\phi_{1},\ldots,\phi_{n})$, the latter allowing for any overdispersion and spatial correlation in the disease data (after the covariate effects have been accounted for). The most common prior for $\bd{\phi}$ is a conditional autoregressive (CAR, \cite{besag1991}) model, which is specified in terms of $n$ univariate conditional distributions, $f(\phi_{k}|\phi_{1},\ldots,\phi_{k-1},\phi_{k+1},\ldots,\phi_{n})$ for $k=1,\ldots,n$. However, each full conditional distribution only depends on the values of $\phi_{j}$ in a small number of \emph{neighbouring} areas. This neighbourhood information is contained in a binary adjacency matrix $W$, which has elements $w_{kj}$ that are equal to one or zero depending on whether areas $(k,j)$ are defined to be neighbours. A common specification is that areas $(k,j)$ are neighbours if they share a common border, which corresponds to $w_{kj}$=1 and is denoted by $k\sim j$. A number of priors have been proposed within the general class of CAR models, and the one we adopt here was originally proposed by \cite{leroux1999}, and has subsequently been reviewed by \cite{macnab2003} and \cite{lee2011}. The model has full conditional distributions given by

\begin{equation}
\phi_{k}|\bd{\phi}_{-k},W,\tau^{2},\rho,\mu\sim\mbox{N}\left(\frac{\rho\sum_{j=1}^{n}w_{kj}\phi_{j} + (1-\rho)\mu}{\rho
\sum_{j=1}^{n}w_{kj} + 1-\rho}~,~\frac{\tau^{2}}{\rho\sum_{j=1}^{n}w_{kj} + 1-\rho}\right).\label{equation conditional phi}
\end{equation}

The conditional expectation is a weighted average of the random effects in neighbouring areas and a global intercept $\mu$ (not included in (\ref{equation diseasemap})), where the weights  are controlled by $\rho$. In this model $\rho=0$ corresponds to independence, while $\rho$ close to one defines strong spatial correlation. These full conditional distributions correspond to a proper multivariate Gaussian distribution if $\rho\in[0,1)$, which is given by

\begin{equation}
\bd{\phi}~\sim~\mbox{N}(\mu\bd{1},~\tau^{2}[\rho W^{*}+(1-\rho)I]^{-1}),\label{equation joint phi}
\end{equation}

where $I$ denotes an $n\times n$ identity matrix while $\bd{1}$ is an n-vector of ones. In the above equation $W^{*}$ has diagonal elements $w_{kk}^{*}=\sum_{i=1}^{n}w_{ki}$, and non-diagonal elements $w_{kj}^{*}=-w_{kj}$. This model simplifies to the intrinsic autoregressive model if $\rho$ is fixed at one, although this does correspond to an improper joint distribution for $\bd{\phi}$.

\subsection{Boundary detection}
\cite{lu2005} propose the use of Boundary Likelihood Values (BLV), which are calculated as $\mbox{BLV}_{kj}=|\hat{R}_{k}-\hat{R}_{j}|$, the absolute difference in risk between two neighbouring areas. The border between neighbouring areas $(k,j)$ can then be classified as a boundary in the risk surface if either

\begin{enumerate}
\item[(a)] $\mbox{BLV}_{kj}>c_{1}$ for some cut-off $c_{1}$; or

\item[(b)] $\mbox{BLV}_{kj}$ is within the top $c_{2}\%$ of the boundary likelihood values over the study region, for some percentage $c_{2}$.
\end{enumerate}

These approaches to boundary detection are ad-hoc, because the decision rules (a) and (b) require tuning constants ($c_{1}$ or $c_{2}$) to be specified.  \cite{jacquez2000} has criticised this approach for this reason, and argues that by specifying the tuning constant the investigator  essentially chooses the number of boundaries that are identified, even though this is unknown and the goal of the analysis.

\section{Methods}
The approach we propose allows the data to determine the number and locations of any boundaries in the risk surface, rather than requiring the investigator to specify a tuning constant. We achieve this by modelling  $w_{kj}$  as a binary random quantity if areas $(k,j)$ share a common border, rather than assuming it is fixed at one.  If $w_{kj}$ is estimated as zero the random effects in areas $(k,j)$ are conditionally independent, which corresponds to a boundary in the risk surface. In contrast, if $w_{kj}$ equals one the random effects are correlated, which corresponds to no boundary. This general approach to boundary detection has previously been proposed by \cite{lu2007}, \cite{ma2007}, and \cite{ma2010}, who model the set of $w_{kj}$ by logistic regression, a CAR prior,  or an Ising model. However, this requires the large set of $w_{kj}$ for all pairs of neighbouring areas  to be estimated, which \cite{li2011} argue are not well identified from the data. In this paper we model the set of $w_{kj}$ as a function of parameters $\bd{\alpha}=(\alpha_{1},\ldots,\alpha_{q})$, rather than treating each $w_{kj}$ as a separate unknown quantity. This results in a parsimonious yet flexible model for detecting boundaries in the risk surface, which avoids the weak parameter identifiability and slow MCMC convergence experienced by \cite{li2011}, when modelling each $w_{kj}$ separately.

\subsection{Level 1 - Observation model}

The first stage of our hierarchical model is similar to that described in Section 2, and is given by

\begin{eqnarray}
Y_{k}|E_{k}, R_{k}&\sim&\mbox{Poisson}(E_{k}R_{k})~~~~\mbox{for }k=1,\ldots,n,\nonumber\\
\ln(R_{k})&=&\phi_{k},\label{equation new1}\\
\phi_{k}|\bd{\phi}_{-k},\mu, \bd{\alpha},\tau^{2}&\sim&\mbox{N}\left(\frac{0.99\sum_{j=1}^{n}w_{kj}(\bd{\alpha})\phi_{j} + 0.01\mu}{0.99\sum_{j=1}^{n}w_{kj}(\bd{\alpha}) + 0.01}~,~\frac{\tau^{2}}{0.99\sum_{j=1}^{n}w_{kj}(\bd{\alpha}) + 0.01}\right).\nonumber
\end{eqnarray}

In common with \cite{lu2007} no covariates are included in (\ref{equation new1}), because $\bd{\phi}$ would then represent the residual pattern in disease risk, and any boundaries identified would hence be in the residual surface. Secondly, we fix $\rho$ at 0.99 because it enforces strong spatial correlation, which allows the presence or absence of boundaries to be determined by $W(\bd{\alpha})$. We note that we do not choose $\rho=1$, as it results in an infinite mean and variance in the conditional distribution of $\phi_{k}$(see (\ref{equation conditional phi})) if an area is surrounded by boundaries, i.e. if $\sum_{j=1}^{n}w_{kj}(\bd{\alpha})=0$ for some area $k$.

\subsection{Level 2 - Neighbourhood model}
We believe that boundaries in the risk surface are likely to occur between populations that are very different, because homogeneous populations should have similar risk profiles. Therefore, we model the presence or absence of a boundary between areas $(k,j)$  by a vector of $q$ non-negative dissimilarity metrics $\mathbf{z}_{kji}=(z_{ki1},\ldots,z_{kjq})$. These metrics have the general form

\begin{equation}
z_{kji}=\frac{|z_{ki}-z_{ji}|}{\sigma_{i}}~~~~\mbox{for }i=1,\ldots,q,\label{equation dissimilarity}
\end{equation}

the absolute difference in the value of a covariate between the two areas in question. Here, $\sigma_{i}$ represents the standard deviation of $|z_{ki}-z_{ji}|$ over all pairs of contiguous areas, and we re-scale the dissimilarity metrics to improve the mixing and convergence of the MCMC algorithm. It is these dissimilarity measures that drive the detection of boundaries in the risk surface, and examples could include differences in the population's social characteristics (e.g. average income) or risk inducing behaviour (e.g. smoking prevalence). Using these metrics we model the  elements of $W(\bd{\alpha})$ as

\begin{eqnarray}
w_{kj}(\bd{\alpha})&=&\left\{ \begin{array}{ll}
1&\mbox{if }\exp(-\sum_{i=1}^{q}z_{kji}\alpha_{i})\geq0.5 \mbox{ and } j\sim k\\
0&\mbox{otherwise}\\\end{array}\right.,\label{equation neighbourhood}
\end{eqnarray}

where pairs of areas that do not share a common border have $w_{kj}(\bd{\alpha})$ fixed at zero.  For areas that are contiguous, the model detects a boundary in the risk surface if\\ $\exp(-\sum_{i=1}^{q}z_{kji}\alpha_{i})$ is less than 0.5. Therefore we constrain the regression parameters to be non-negative, so that the greater the dissimilarity between two areas the more likely there is to be a boundary between them. In contrast, if two areas have identical covariate values (and hence homogeneous populations) there cannot be a boundary between them, regardless of the value of $\bd{\alpha}$. This is the reason we do not include an intercept term in (\ref{equation neighbourhood}), as doing so would allow boundaries to be detected between areas with homogeneous populations. The regression parameters determine the number of risk boundaries in the study region, with larger values of $\bd{\alpha}$ corresponding to more boundaries being detected. If only one dissimilarity metric $z$ is included in (\ref{equation neighbourhood}), then a plausible range of values for the single regression parameter $\alpha$ can be determined. At one extreme, no boundaries will be detected if $\alpha\leq-\ln(0.5)/z^{max}$, while at the other, all borders in the study region will be considered as boundaries (unless $z_{kj}$=0) if $\alpha>-\ln(0.5)/z^{min}$. Here $(z^{min}, z^{max})$ denote the minimum positive and maximum values of the dissimilarity metric. More generally, if there are $q$ dissimilarity metrics then boundaries are identified if

$$\exp(-z_{kj1}\alpha_{1})\times\ldots\times\exp(-z_{kjq}\alpha_{q})<0.5,$$

where the value of each component, $\exp(-z_{kji}\alpha_{i})$, must lie between zero and  one. Therefore, if $\alpha_{i}\leq-\ln(0.5)/z_{i}^{max}$, the dissimilarity measure  $z_{kji}$ is not solely responsible for detecting any boundaries, because $\exp(-z_{i}\alpha_{i})$ would be greater than 0.5 for all pairs of contiguous areas. Therefore in terms of interpretation, the dissimilarity metric can be said to have no effect on detecting boundaries if the entire 95$\%$ credible interval for $\alpha_{i}$ is less than $\alpha_{min}=-\ln(0.5)/z_{kji}^{max}$. In contrast, if the interval lies completely above $\alpha_{min}$, then the metric can be said to have a substantial effect on identifying risk boundaries. We note that the usual statistical representation of `no effect' (credible interval that includes zero) is not possible in this context, because the regression parameters are constrained to be non-negative. We also note that the approach we outline does not guarantee that the boundaries we detect will be closed (form an unbroken line, an example of which is shown in Figure \ref{figure results}), which allows us to detect boundaries that enclose an entire subregion, as well as those that just separate highly different areas.

\subsection{Level 3 - Hyperpriors}
The vector of random effects depend on hyperparameters $(\mu, \tau^{2},\bd{\alpha})$, which respectively control its mean, variance and correlation structure. The mean $\mu$ is assigned a weakly informative Gaussian prior distribution, with a mean of  zero and a variance of 10. The variance parameter $\tau^{2}$ is assigned a weakly informative Uniform$(0,10)$ prior on the standard deviation scale, following the suggestion of \cite{gelman2006}. We adopt a uniform prior for the regression parameters, $\alpha_{i}\sim\mbox{Uniform}(0,M_{i})$, which corresponds to our prior ignorance about the number of boundaries in the risk surface. An alternative would be a reciprocal prior, $f(\alpha_{i})\propto\frac{1}{\alpha_{i}}I[0\leq \alpha_{i}\leq M_{i}]$, which represents our prior belief that the risk surface is spatially smooth. In both cases a natural upper limit would be $M_{i}=-\ln(0.5)/z_{kji}^{min}$, the value at which the dissimilarity measure $z_{kji}$ solely identifies all borders as boundaries in the risk surface. However, in a boundary detection analysis one is looking to identify boundaries between collections of areas, which have similar risks within each collection but differ across the boundary. Therefore we fix $M_{i}$ so that at most 50$\%$ of borders can be classified as boundaries, and present a sensitivity analysis in the supplementary material to this choice of 50$\%$.

\section{Simulation study}
In this section we present a simulation study, that assesses the accuracy with which our proposed model can detect boundaries in the risk surface. In doing this we assess whether our model can detect `true' boundaries in the risk surface, as well as the extent to which it falsely identifies boundaries that do not exist. We have decided not to compare our model to the existing boundary detection approach using (\ref{equation diseasemap}), (\ref{equation conditional phi}) and BLVs, because this requires a tuning constant ($c_{1}$ or $c_{2}$) to be specified by the user, which in real life would be unknown. However, these tuning constants are known in this simulation setting, and specifying their value would put this method at an unfair advantage or disadvantage, depending on whether we chose the `correct' values.\\

\subsection{Data generation}
\begin{figure}
\centering\caption{Locations of the true boundaries in the simulated risk surfaces.}
\label{figure boundaries}\scalebox{0.4}{\includegraphics{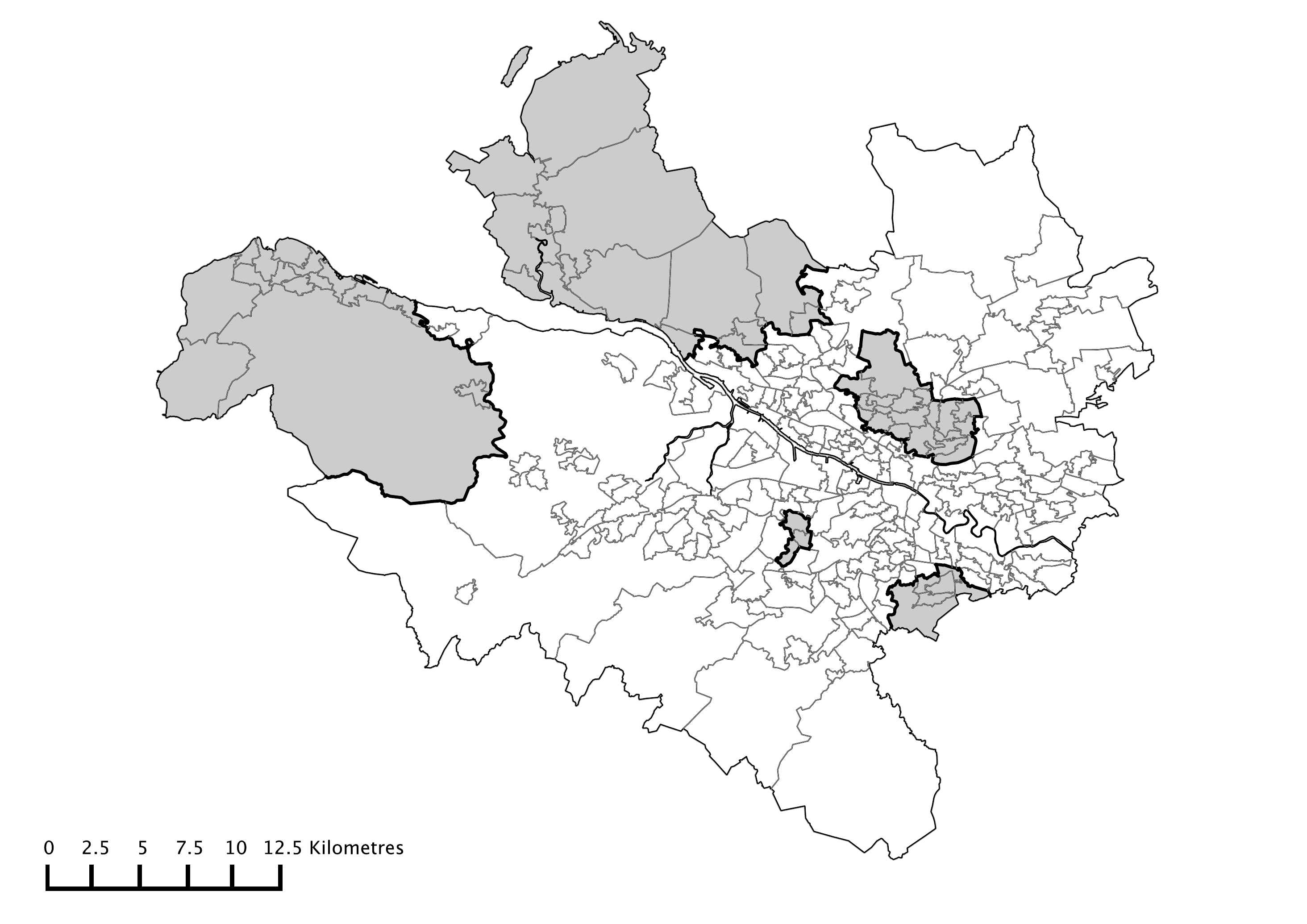}}
\end{figure}

We base our study on the $n=271$ areas that comprise the Greater Glasgow and Clyde health board, which is the region considered in the cancer mapping study presented in Section five. Disease counts are generated from the Poisson model (\ref{equation new1}), where the expected numbers of admissions, $\mathbf{E}$, relate to the Glasgow cancer data. A new risk surface $\bd{R}=\exp(\bd{\phi})$ is generated for each set of simulated disease data, because this ensures the results are not affected by a particular realisation of $\bd{\phi}$. Each simulated risk surface has fixed boundaries, which are shown by the bold black lines in Figure \ref{figure boundaries}. There are 74 boundaries in total, which corresponds to approximately 10$\%$ of the set of borders in the study region. This set of boundaries partition the study region into 6 groups, the main area shaded in white, and the remaining 5 smaller areas shaded in grey. To produce risk surfaces with boundaries, the random effects ($\bd{\phi}$) are generated from a multivariate Gaussian distribution with a piecewise constant mean, which in the white region is equal to 0, while in the grey regions it is equal to $k_{1}$.  The correlation function is from the Matern class with smoothness parameter equal to $\kappa=2.5$, while the spatial range is fixed so that the median correlation between areas is 0.5. The dissimilarity metrics are generated from

\begin{equation}
z_{kj}\sim\left\{\begin{array}{cc}|\mbox{N}(1,0.5^{2})|&\mbox{   if areas $(k,j)$ are not separated by a boundary.}\\
|\mbox{N}(1+k_{2},0.5^{2})|&\mbox{   if areas $(k,j)$ are separated by a boundary.}\end{array}\right.\label{equation simstudy}
\end{equation}

Here, larger values of $k_{2}$ correspond to dissimilarity metrics that better identify the true boundaries in the risk surface; i.e. have larger values for the boundaries in Figure \ref{figure boundaries} than for the non-boundaries. We assess the effect on model performance of changing both  the size of the boundaries (via $k_{1}$) and the quality of the dissimilarity metrics (via $k_{2}$), the results of which are displayed in Table \ref{table simulation}. When assessing the effect of boundary size we fix $k_{2}=3$, which provides nearly `perfect' dissimilarity metrics, i.e. values for $z_{kj}$ for boundaries and non-boundaries that almost never overlap. Conversely, when assessing the effect of having imperfect dissimilarity metrics, we fix the boundary size at $k_{1}=0.4$, which provides fairly large boundaries to identify.

\subsection{Results}
The top half of Table \ref{table simulation} displays the effects of changing the magnitude of the risk boundaries, in the idealised situation of having perfect dissimilarity metrics. The constant $k_{1}$ represents the difference in the mean value of the risk surface between white and grey areas (see Figure \ref{figure boundaries}), hence smaller values correspond to a spatially smoother surface ($k_{1}=0$ would correspond to a completely smooth risk surface with no boundaries). The table shows that the model can detect larger risk boundaries more often than smaller ones as expected, although it still achieves over a 90$\%$ detection rate when the risk surface only has a mean difference of 0.2. The much lower percentages for smaller values of $k_1$ are also not surprising, as they correspond to situations in which the average size of the true boundaries are not very different to the average size of the non-boundaries. In addition, the false positive rates are very low (generally less than $1\%$) regardless of the boundary size, which suggests that detected boundaries are likely to be real.\\

The bottom half of Table \ref{table simulation} displays the effects of having imperfect dissimilarity metrics, i.e. metrics for which some of the values corresponding to the `true' boundaries  are smaller than some of those corresponding to the non-boundaries. As the constant $k_{2}$ decreases, there is a greater overlap between the values of the dissimilarity metrics at boundaries and non-boundaries. The limit of $k_{2}=0$ corresponds to a dissimilarity metric with no information, i.e. it has the same range of values for boundaries as non-boundaries. The table shows that if the dissimilarity metric is nearly perfect ($k_2=3$ corresponds to the means in equation (\ref{equation simstudy}) being separated by 6 standard deviations), then the model nearly always correctly identifies boundaries and non-boundaries. However, as the information content in the dissimilarity metric decreases (as $k_2$ decreases) so does the performance of the model, both in terms of the boundary agreement and the false positive rate. If the dissimilarity metric contains no information the model only identifies 1.86$\%$ of the true boundaries as would be expected, although in this situation the false positive rate returns to being low at around 1$\%$.

\begin{table}
\caption{\label{table simulation} The effect of boundary size (as measured by $k_{1}$) and the quality of the dissimilarity metrics (as measured by $k_{2}$) on the effectiveness of the model. The table displays the percentage agreement for boundaries (BA) and non-boundaries (NBA), as well as the bias and root mean square error (RMSE)  of the estimated risk surface, which are presented as a percentage of their true value.}
\centering\begin{tabular}{cccrrrr}

\hline
\textbf{Comparison}&\textbf{$k_{1}$} &$k_{2}$& \textbf{BA ($\%$)} & \textbf{NBA ($\%$)}&\textbf{Bias} & \textbf{RMSE} \\  \hline

&0.4&3&99.97&98.70&-0.123&5.689\\
Boundary&0.3&3&99.57&99.16&-0.195&5.700\\
Size&0.2&3&93.76&99.43&-0.187&5.689\\
&0.1&3&48.31&99.89&-0.092&5.706\\
&0.05&3&25.84&100&-0.139&5.663\\\hline

&0.4&3&99.97&98.70&-0.123&5.689\\
Quality&0.4&2&98.89&95.07&-0.134&5.747\\
of $z_{kj}$&0.4&1.5&96.27&88.37&-0.140&5.866\\
&0.4&1&87.19&80.85&-0.206&6.226\\
&0.4&0.5&55.74&80.93&-0.242&6.927\\
&0.4&0&1.85&98.82&-0.248&7.178\\\hline
\end{tabular}
\end{table}

\section{Case study - Cancer risk in Greater Glasgow}
This section presents a study mapping the risk of lung cancer in Greater Glasgow, Scotland, between 2001 and 2005.

\subsection{Data description}
The data for our study are publicly available, and can be downloaded from the Scottish Neighbourhood Statistics (SNS) database (\emph{http://www.sns.gov.uk}). The study region is the Greater Glasgow and Clyde health board, which contains the city of Glasgow in the east, and the river Clyde estuary in the west. Glasgow is known to contain some of the poorest people in Europe (\cite{leyland2007}), and has rich and poor communities that are geographically adjacent. The Greater Glasgow and Clyde health board is partitioned into $n=271$ administrative units called Intermediate Geographies (IG), which were developed specifically for the distribution of small-area statistics, and have a median area of 124 hectares and a median population of 4,239.\\

\begin{figure}
\centering\caption{Standardised Incidence Ratio (SIR) for lung cancer in Greater Glasgow between 2001 and 2005.}
\label{figure SIR}\scalebox{0.4}{\includegraphics{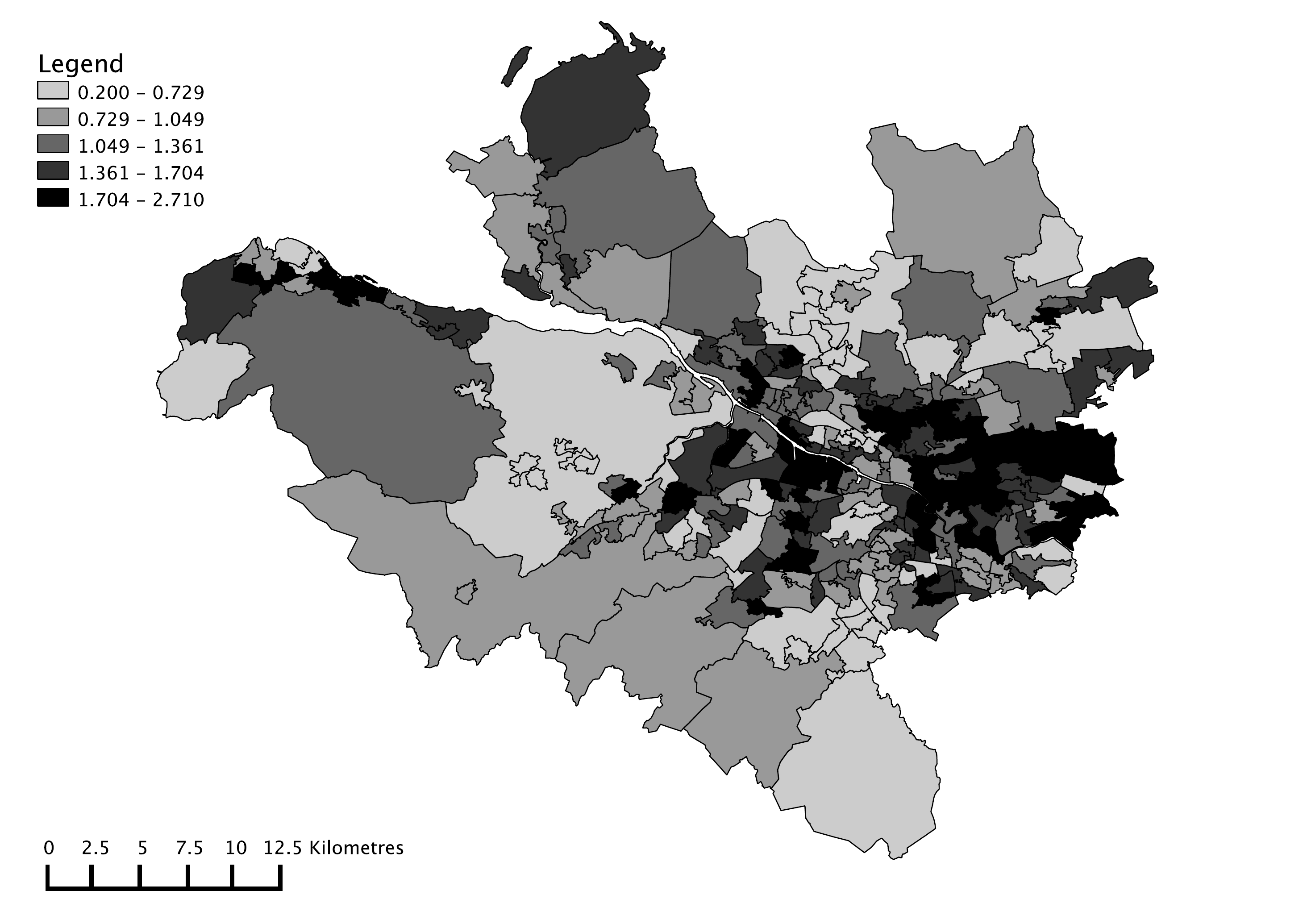}}
\end{figure}

The disease data we model are the number of people diagnosed with lung cancer between 2001 and 2005 in each IG, which corresponds to  ICD-10 codes C33 - C34. The expected numbers of cases in each IG are calculated by external standardisation, using age and sex adjusted rates for the whole of Scotland. These rates were obtained from the Information Services Division (ISD), which is the statistical arm of the National Health Service in Scotland. The simplest measure of disease risk is the standardised incidence ratio, which is presented in Figure \ref{figure SIR} as a choropleth map.  The Figure shows that the risk of lung cancer is highest in the heavily deprived east end of Glasgow (east of the study region), as well as along the banks of the river Clyde (the thin white line running south east). The Figure also shows that cancer incidence in Greater Glasgow is higher than in the rest of Scotland, as the average SIR across the study region is 1.186.\\

Large amounts of covariate data are available from the Scottish Neighbourhood Statistics database, and the first variable we consider is a modelled estimate of the percentage of the population in each IG that smoke, further details of which are available from \cite{smoke}. The causal relationship between smoking and lung cancer risk is long standing (see for example \cite{doll1950} and \cite{doll2005}), and it is likely to be the most important dissimilarity metric in our study. Cancer risk has also been shown to vary by ethnic group (see for example \cite{NCIN}), so we include the percentage of school children from ethnic minorities as a proxy measure. Socio-economic deprivation is also associated with cancer risk (see for example \cite{quinn2000} and \cite{woods2006}), and as a proxy measure we use the natural log of the median house price. This proxy measure is used because it is the measure of deprivation available that is least correlated with the smoking covariate (Pearson's r=-0.69). Finally, we acknowledge that other factors such as diet and physical activity have been repeatedly associated with lung cancer risk. However, no data are available at the small-area level about these factors.

\subsection{Results - modelling}
Inference for all models is based on 50,000 MCMC samples generated from five Markov chains, that were initialised at dispersed locations in the sample space. Each chain is burnt-in until convergence (40,000 iterations), and the next 10,000 samples are used for the analysis. A number of models were fitted to the data with different combinations  of the three dissimilarity metrics, and in each case the residuals were assessed for the presence of spatial correlation. This was achieved using a permutation test based on Moran's I statistic (\cite{moran1950}), and in all cases the model adequately removes the spatial correlation present in the data, as the corresponding p-values (not shown) are greater than 0.05.\\

 Initially, all three covariates (smoking, ethnicity and house price) were included in the model as dissimilarity metrics, and the posterior medians and 95$\%$ credible intervals are displayed in Table \ref{table results1}. Also displayed in Table \ref{table results1} is $\alpha_{min}$, the threshold value below which the dissimilarity metric does not solely detect any boundaries in the risk surface. The table shows that smoking prevalence has substantial influence in detecting boundaries, as the estimate and credible interval lie above the threshold value of $\alpha_{min}=0.131$. In contrast, neither ethnicity nor house price have any effects in detecting risk boundaries, as their credible intervals both lie below the corresponding `no effect' thresholds. This suggests that the existence of risk boundaries only  depend on smoking prevalence, and that the other covariates do not add anything to the model.\\

To provide more evidence for this, the fit of the four possible models that include smoking prevalence as a dissimilarity metric were compared, which included the full model, smoking on its own, and smoking with each of the two remaining covariates separately. In all cases both the Deviance Information Criterion and the  number of boundaries detected remained largely unchanged. In the smoking only model the sole regression parameter has a posterior median and 95$\%$ credible interval of  0.257 (0.239, 0.262), which, in common with the results in Table \ref{table results1}, lies completely above the `no effect' threshold.

\begin{table}
\caption{\label{table results1} Covariate effects from the initial model, presented as estimates (posterior medians) and 95$\%$ credible intervals. In addition, the threshold value $\alpha_{min}=-\ln(0.5)/z^{max}$ is presented, the value at which each dissimilarity metric does not solely identify any risk boundaries.}
\centering\begin{tabular}{crrr}

\hline
\textbf{Covariate}&\textbf{Estimate} & \textbf{95$\%$ credible interval} & \textbf{$\alpha_{min}$} \\  \hline

Smoking&0.232&(0.171, 0.254)&0.131\\
Ethnicity&0.012&(0.001, 0.046)&0.126\\
House price (log)&0.015&(0.001, 0.103)&0.119\\
\hline
\end{tabular}
\end{table}

\subsection{Results - risk maps}
\begin{figure}
\centering\caption{Estimated risk surface for lung cancer in Greater Glasgow between 2001 and 2005. The risk boundaries are denoted by solid white lines.}
\label{figure results}\scalebox{0.4}{\includegraphics{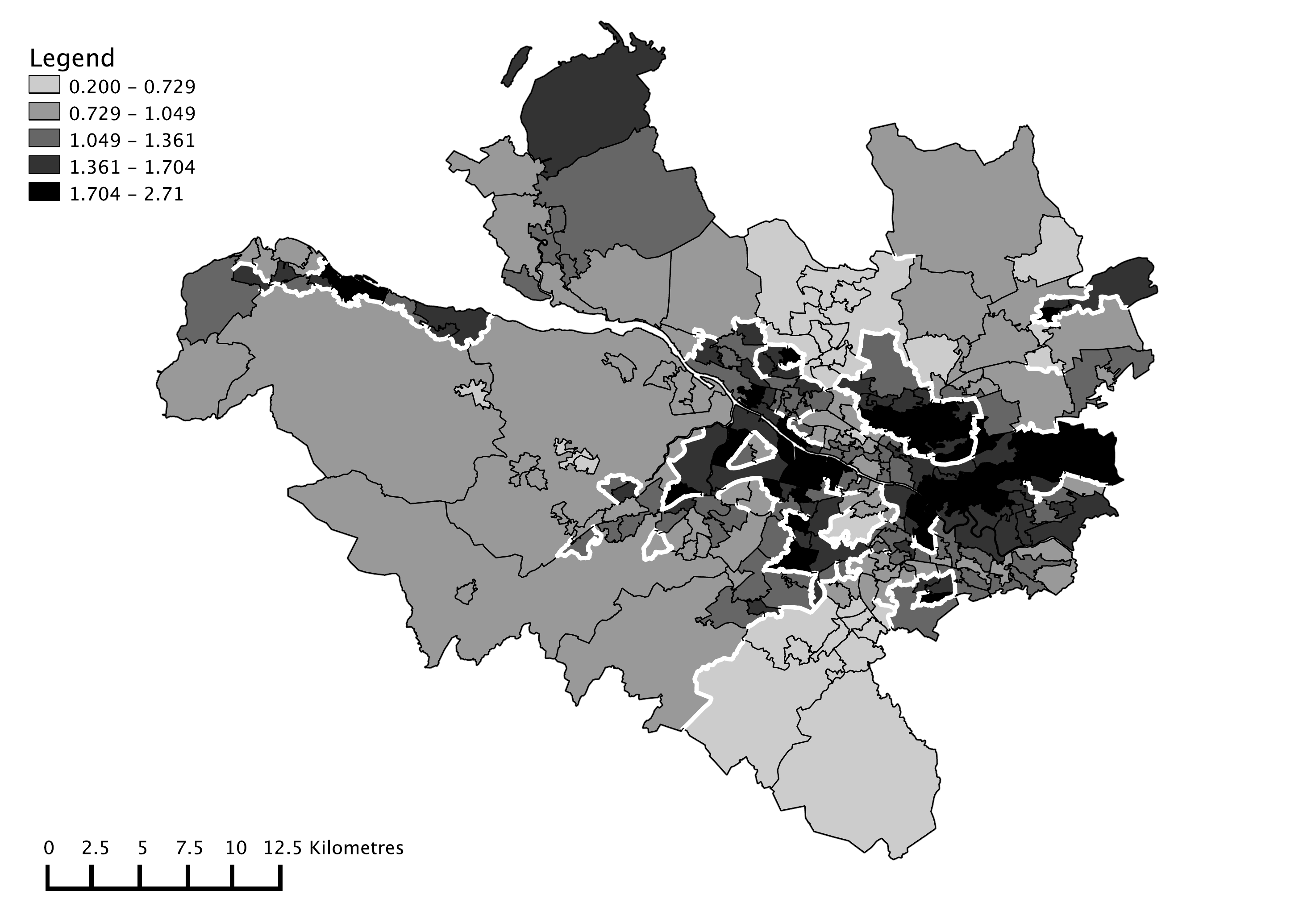}}
\end{figure}

Figure \ref{figure results} displays the estimated risk surface for lung cancer from the smoking only model, where the shading is on the same scale as that used in Figure \ref{figure SIR}. The solid white lines denote the risk boundaries that have been identified by the smoking covariate, which are defined by having posterior median values of $w_{kj}$ equal to zero. We note that the study region is split completely in two by the river Clyde (the thin white line running south east), and areas on opposite banks are not assumed to be neighbours. Therefore no boundaries can be detected across the river, which explains the absence of boundaries in this area.\\

The figure shows that the smoking covariate detects 162 boundaries in the risk surface (23.1$\%$ of all possible borders), including within the city of Glasgow (middle of the map) as well as along the southern coast of the Clyde estuary (far west of the map). The majority of these estimated risk boundaries appear to correspond to sizeable changes in the risk surface, suggesting that the smoking covariate appears to be an appropriate dissimilarity metric for detecting such boundaries. However,  a few of the boundaries identified show no evidence of separating areas with differing health risks, such as the closed boundary in the south of the city. These `false positives' correspond to two areas having different smoking prevalences but similar risk profiles, and would be a starting point for a more detailed investigation into why the risk profiles are similar given the vastly different smoking rates.

\section{Discussion}
In this paper we have proposed a statistical approach to detecting boundaries in disease risk maps, which separate populations that exhibit high and low risks of disease. Our approach detects boundaries by measuring the dissimilarity between populations living in neighbouring areas, because we believe that abrupt changes in the risk surface are most likely to occur between populations that are geographically adjacent but have very different social characteristics or risk inducing behaviour. Our approach has the advantage of being fully automatic, so that unlike the approach of \cite{lu2005}, the number of boundaries in the risk surface is determined by the data and not \emph{a-priori} by the investigator. In addition, our approach identifies boundaries using a small number of regression parameters $\bd{\alpha}$, rather than estimating the set of neighbourhood relations $\{w_{kj}\}$ for all pairs of contiguous areas. \cite{li2011} have suggested that this latter approach (used by \cite{lu2007} and \cite{ma2007}) can lead to identifiability problems and slow MCMC convergence, due to the large number of parameters to be estimated. For example, in the lung cancer application presented in Section 5, there are 701 neighbourhood relations $w_{kj}$, compared with disease data in only 271 areas.\\

The simulation study presented in Section 4 suggests that our approach generally performs well, both in terms of detecting true boundaries in the risk surface, as well as not detecting large numbers of false positives. In the presence of a perfect dissimilarity metric our approach can detect the majority of true risk boundaries, for example, it has a 99.6$\%$ detection rate when the average difference in the log risk surface is 0.3. The main drawback of our approach is illustrated by the bottom half of Table \ref{table simulation}, which shows that it is crucially dependent on the existence of good quality dissimilarity metrics, that have large values for risk boundaries and small values for non-boundaries. However, in the absence of good dissimilarity metrics (as in the last few rows of Table \ref{table simulation}) our approach appears to remain conservative,  in the sense that the false positive rate is relatively low. \\

As our approach to boundary detection is based solely on covariates, it has the advantage that in addition to identifying risk boundaries, it also identifies the underlying drivers of these boundaries.  However, as illustrated by the simulation study, the corresponding disadvantage is that it relies on the existence of relevant covariate data, which may not always be available. Our lung cancer example also illustrates this, as the main covariate (smoking prevalence) identifies the majority of the risk boundaries evident from Figure \ref{figure results}. However, it also identifies some boundaries that do not appear to be real, as well as ignoring others where there is a suspected discontinuity in the risk surface. This imperfection could be caused by the omission of other important dissimilarity metrics, or by the fact that the smoking covariate is a modelled estimate rather than being real prevalence data. Therefore, the production of risk maps such as Figure \ref{figure results} can be viewed as an exploratory tool, which helps the investigator understand the drivers of the spatial variation in disease risk. For example, if a risk boundary is not detected despite their being a discontinuity in the risk surface, further investigation of the two areas in question could be carried out to determine what other factors could be causing this discontinuity.\\

Finally, this paper opens up numerous avenues for future work. The most obvious is to develop a boundary detection method that has the advantages of the method proposed here, but does not rely on covariate data to detect risk boundaries. One possibility in this vein is to develop an iterative algorithm, which compares the fit (possibly using DIC) of a number of models with different but fixed neighbourhood matrices $W$. In this way one could compare a model where all adjacent areas have $w_{kj}=1$, against an alternative with $w_{kj}=0$ for areas that are suspected of being separated by a discontinuity in the risk surface. Other natural extensions to this approach include the detection of boundaries in multiple disease risk surfaces simultaneously, as well as adding a temporal dimension to the model.

\section*{Acknowledgments}
The data and shapefiles used in this study were provided by the  Scottish Government. \emph{Conflict of Interest:} None declared.

\section*{Funding}
This work was supported by the Economic and Social Research Council [RES-000-22-4256].

\bibliographystyle{chicago}
\bibliography{literature}

\begin{thebibliography}{}

\bibitem[\protect\citeauthoryear{Banerjee, Carlin, and Gelfand}{Banerjee
  et~al.}{2004}]{banerjee2004}
Banerjee, S., B.~Carlin, and A.~Gelfand (2004).
\newblock {\em Hierarchical {M}odelling and {A}nalysis for {S}patial {D}ata\/}
  (1st ed.).
\newblock Chapman and {H}all.

\bibitem[\protect\citeauthoryear{Besag, York, and Mollie}{Besag
  et~al.}{1991}]{besag1991}
Besag, J., J.~York, and A.~Mollie (1991).
\newblock Bayesian image restoration with two applications in spatial
  statistics.
\newblock {\em Annals of the {I}nstitute of {S}tatistics and
  {M}athematics\/}~{\em 43}, 1--59.

\bibitem[\protect\citeauthoryear{Boots}{Boots}{2001}]{boots2001}
Boots, B. (2001).
\newblock Using local statistics for boundary characterization.
\newblock {\em Geo{J}ournal\/}~{\em 53}, 339--345.

\bibitem[\protect\citeauthoryear{Doll and {Bradford Hill}}{Doll and {Bradford
  Hill}}{1950}]{doll1950}
Doll, R. and A.~{Bradford Hill} (1950).
\newblock Smoking and carcinoma of the lung.
\newblock {\em British {M}edical {J}ournal\/}~{\em 2}, 739Ð748.

\bibitem[\protect\citeauthoryear{Doll, Peto, Boreham, and Sutherland}{Doll
  et~al.}{2005}]{doll2005}
Doll, R., R.~Peto, J.~Boreham, and I.~Sutherland (2005).
\newblock Mortality from cancer in relation to smoking: 50 years observations
  on british doctors.
\newblock {\em British {J}ournal of {C}ancer\/}~{\em 92}, 426--429.

\bibitem[\protect\citeauthoryear{Elliott, Wakefield, Best, and Briggs}{Elliott
  et~al.}{2000}]{elliott2000}
Elliott, P., J.~Wakefield, N.~Best, and D.~Briggs (2000).
\newblock {\em Spatial {E}pidemiology: {M}ethods and {A}pplications\/} (1st
  ed.).
\newblock Oxford {U}niversity {P}ress.

\bibitem[\protect\citeauthoryear{Gelman}{Gelman}{2006}]{gelman2006}
Gelman, A. (2006).
\newblock Prior distributions for variance parameters in hierarchical models.
\newblock {\em Bayesian {A}nalysis\/}~{\em 1}, 515--533.

\bibitem[\protect\citeauthoryear{Jacquez, Maruca, and Fortin}{Jacquez
  et~al.}{2000}]{jacquez2000}
Jacquez, G., S.~Maruca, and M.~Fortin (2000).
\newblock From fields to objects: {A} review of geographic boundary analysis.
\newblock {\em Journal of {G}eographical {S}ystems\/}~{\em 2}, 221--241.

\bibitem[\protect\citeauthoryear{Lawson}{Lawson}{2008}]{lawson2008}
Lawson, A. (2008).
\newblock {\em Bayesian {D}isease {M}apping: {H}ierarchical {M}odelling in
  {S}patial {E}pidemiology\/} (1st ed.).
\newblock Chapman and {H}all.

\bibitem[\protect\citeauthoryear{Lee}{Lee}{2011}]{lee2011}
Lee, D. (2011).
\newblock A comparison of conditional autoregressive model used in {B}ayesian
  disease mapping.
\newblock {\em Spatial and {S}patio-temporal {E}pidemiology\/}~{\em to appear},
  DOI:10.1016/j.sste.2011.03.001.

\bibitem[\protect\citeauthoryear{Leroux, Lei, and Breslow}{Leroux
  et~al.}{1999}]{leroux1999}
Leroux, B., X.~Lei, and N.~Breslow (1999).
\newblock {\em Estimation of disease rates in small areas: {A} new mixed model
  for spatial dependence}, Chapter Statistical {M}odels in {E}pidemiology, the
  {E}nvironment and {C}linical {T}rials, Halloran, M and Berry, D (eds), pp.\
  135--178.
\newblock Springer-Verlag, {N}ew {Y}ork.

\bibitem[\protect\citeauthoryear{Leyland, Dundas, McLoone, and Boddy}{Leyland
  et~al.}{2007}]{leyland2007}
Leyland, A., R.~Dundas, P.~McLoone, and F.~Boddy (2007).
\newblock Inequalities in mortality in {S}cotland 1981-2001.
\newblock {\em BMC Public Health\/}~{\em 7}, 172.

\bibitem[\protect\citeauthoryear{Li, banerjee, and Mc{B}ean}{Li
  et~al.}{2011}]{li2011}
Li, P., S.~banerjee, and A.~Mc{B}ean (2011).
\newblock Mining boundary effects in areally referenced spatial data using the
  {B}ayesian information criterion.
\newblock {\em Geoinformatica\/}~{\em to appear}, DOI:
  10.1007/s10707--010--0109--0.

\bibitem[\protect\citeauthoryear{Lu and Carlin}{Lu and Carlin}{2005}]{lu2005}
Lu, H. and B.~Carlin (2005).
\newblock Bayesian {A}real {W}ombling for {G}eographical {B}oundary {A}nalysis.
\newblock {\em Geographical {A}nalysis\/}~{\em 37}, 265--285.

\bibitem[\protect\citeauthoryear{Lu, Reilly, Banerjee, and Carlin}{Lu
  et~al.}{2007}]{lu2007}
Lu, H., C.~Reilly, S.~Banerjee, and B.~Carlin (2007).
\newblock Bayesian areal wombling via adjacency modelling.
\newblock {\em Environmental and {E}cological {S}tatistics\/}~{\em 14},
  433--452.

\bibitem[\protect\citeauthoryear{Ma and Carlin}{Ma and Carlin}{2007}]{ma2007}
Ma, H. and B.~Carlin (2007).
\newblock Bayesian {M}ultivariate {A}real {W}ombling for {M}ultiple {D}isease
  {B}oundary {A}nalysis.
\newblock {\em Bayesian {A}nalysis\/}~{\em 2}, 281--302.

\bibitem[\protect\citeauthoryear{Ma, Carlin, and Banerjee}{Ma
  et~al.}{2010}]{ma2010}
Ma, H., B.~Carlin, and S.~Banerjee (2010).
\newblock Hierarchical and {J}oint {S}ite-{E}dge {M}ethods for {M}edicare
  {H}ospice {S}ervice {R}egion {B}oundary {A}nalysis.
\newblock {\em Biometrics\/}~{\em 66}, 355--364.

\bibitem[\protect\citeauthoryear{MacNab}{MacNab}{2003}]{macnab2003}
MacNab, Y. (2003).
\newblock Hierarchical {B}ayesian {M}odelling of {S}patially {C}orrelated
  {H}ealth {S}ervice {O}utcome and {U}tilization {R}ates.
\newblock {\em Biometrics\/}~{\em 59}, 305--316.

\bibitem[\protect\citeauthoryear{Moran}{Moran}{1950}]{moran1950}
Moran, P. (1950).
\newblock Notes on continuous stochastic phenomena.
\newblock {\em Biometrika\/}~{\em 37}, 17--23.

\bibitem[\protect\citeauthoryear{{National {C}ancer {I}ntelligence
  {N}etwork}}{{National {C}ancer {I}ntelligence {N}etwork}}{2009}]{NCIN}
{National {C}ancer {I}ntelligence {N}etwork} (2009).
\newblock Cancer incidence and survival by major ethnic group, england,
  2002-2006.
\newblock Technical report, Cancer Research UK Cancer Survival Group, and
  London School of Hygiene and Tropical Medicine.

\bibitem[\protect\citeauthoryear{{Quinn, M and Babb, P}}{{Quinn, M and Babb,
  P}}{2000}]{quinn2000}
{Quinn, M and Babb, P} (2000).
\newblock Cancer trends in {E}ngland and {W}ales, 1950Ð1999.
\newblock Technical report, Health {S}tatistics {Q}uarterly, {O}ffice for
  {N}ational {S}tatistics.

\bibitem[\protect\citeauthoryear{Wakefield}{Wakefield}{2007}]{wakefield2007}
Wakefield, J. (2007).
\newblock Disease mapping and spatial regression with count data.
\newblock {\em Biostatistics\/}~{\em 8}, 158--183.

\bibitem[\protect\citeauthoryear{Whyte, Gordon, Haw, Fischbacher, and
  Harrison}{Whyte et~al.}{2007}]{smoke}
Whyte, B., D.~Gordon, S.~Haw, C.~Fischbacher, and R.~Harrison (2007).
\newblock {\em An atlas of tobacco smoking in Scotland: A report presenting
  estimated smoking prevalence and smoking attributable deaths within
  Scotland}.
\newblock NHS Health Scotland.

\bibitem[\protect\citeauthoryear{Womble}{Womble}{1951}]{womble1951}
Womble, W. (1951).
\newblock Differential systematics.
\newblock {\em Science\/}~{\em 114}, 315--322.

\bibitem[\protect\citeauthoryear{Woods, Rachet, and Coleman}{Woods
  et~al.}{2006}]{woods2006}
Woods, L., B.~Rachet, and M.~Coleman (2006).
\newblock Origins of socio-economic inequalities in cancer survival: a review.
\newblock {\em Annals of {O}ncology\/}~{\em 17}, 5--19.

\end{thebibliography}

\end{document}